\documentstyle[psfig,12pt]{article}
\def\reference{\parskip 0pt\par\noindent\hangindent 0.5 truecm}

\begin{document}

\title{High Energy Radiation Generated at Boundary Shear Layers of Relativistic Jets }

\author{ \L . Stawarz$^{1}$ \& 
 M. Ostrowski
} 

\date{}

\maketitle

{\center
Obserwatorium Astronomiczne, Uniwersytet Jagiello\'{n}ski, ul. Orla 171, Krak\'{o}w, Poland\\[3mm]
$^1$ stawarz@oa.uj.edu.pl\\[3mm]
}

\begin{abstract}

A simple model of cosmic ray electron acceleration at the jet boundary (Ostrowski 2000) yields a power-law particle energy distribution of ultrarelativistic electrons with an energy cut-off growing with time and finally, a growing particle bump at the energy, where energy gains equal radiation losses. For such electron distribution, in tens-of-kpc scale jets, we derived the observed time varying spectra of synchrotron and inverse-Compton radiation, including comptonization of synchrotron and cosmic microwave background photons. Slowly varying spectral index along the jet in the `low-frequency' spectral range is a natural consequence of boundary layer acceleration. Variations of the high energy bump of the electron distribution can give rise to anomalous behaviour in X-ray band in comparison to the lower frequencies. 

\end{abstract}


\bigskip

\section{Introduction}

Particle acceleration in a velocity shear layer was proposed in the early eighties by Berezhko with collaborators and then gradually developed (cf. a review by Berezhko 1990). Action of such a mechanism at relativistic shear layers occurring at side boundaries of relativistic jets was discussed by Ostrowski (1990, 1998, 2000). The mechanism was considered to provide ultra high energy cosmic rays and to be an important factor influencing the dynamics of relativistic jets in extragalactic radio sources thereby forming {\it cosmic rays cocoons}. Ostrowski \& Sikora (2001) proposed that such acceleration mechanism can provide a cosmic ray proton population being a substantial pressure factor in FR II radio source lobes. 

The existence of the boundary layer with a velocity shear was first suggested by the observations of radio and optical jets in FR I sources. Owen et al. (1989) concluded that the jet morphology in M87 is dominated by the boundary layer 
more than by the shock structure. Constancy of radio-to-optical spectral index along the jet and similarity of radio and optical maps with emissivity peaking near the jet's surface strongly suggested acceleration of radiating 
electrons {\it in situ} at the boundary layer. Later radio and optical polarimetry of M87 (Perlman et al. 1999) confirmed the complex structure of its jet containing a shear layer with a highly ordered magnetic field parallel to the jets axis. A map of the radio spectral index of another well known BL Lac, Mkn 501, revealed a boundary region of its jet with an unexpectedly flat spectrum (Edwards et al. 2000), suggesting action of some reacceleration mechanism at the jet surface.

Some FR II radio sources also reveal {\it spine - shear layer} jet morphology. For instance radio observations of 3C353 (Swain et al. 1998) was interpreted in terms of a Doppler-hidden relativistic spine surrounded by the slower moving 
boundary region. VLBI observations of another FR II object 1055+018 (Attridge et al. 1999) revealed a fragmentary but distinct boundary layer with the mean longitudinal magnetic field present where interactions with surrounding matter were strongest (i.e., where the jet bended). Optical and radio observations of the quasar 3C 273 (Jester et al. 2001) showed smooth changes of the radio-to-optical spectral index along the jet. This strongly suggested reacceleration of the radiating electrons taking place not in the specific shock-like regions, but within the {\it whole} jet.    

One should note that our knowledge of physical conditions at the relativistic jet boundary is still insufficient. Cosmic ray viscosity and radiation viscosity in a boundary layer with a velocity shear influencing hydrodynamics of the jet and 
particle distribution along its surface was studied by Earl et al. (1988), Jokipii et al. (1989) and Arav \& Begelman (1992), respectively. On the other hand, 3D simulations revealed the presence of a jet shear layer with a high specific internal energy and highly turbulent, subsonic and thin cocoon surrounding the relativistic flow (Aloy et al. 1999). Simulations of radio emission from such jets (Aloy et al. 2000) show radial structures in both intensity and polarization.

\section{Acceleration of Cosmic Ray Electrons} 

Let us assume that a thickness of the transition layer between the jet and the surrounding medium, $D$, is comparable to the jet radius $R_j \approx 1 \, {\rm kpc}$ (cf. Owen et al. 1989, Swain et al. 1998), and that the magnetic field ${\bf B}$ frozen into tenuous plasma at and outside the boundary is on average parallel to the flow velocity ${\bf U}$. Acceleration of energetic particles injected into the boundary layer results from its scattering on irregularities of the magnetic field in a
turbulent medium with a perpendicular to ${\bf U}$ mean velocity gradient. In the considered case of the mean magnetic field aligned along the jet in a highly turbulent boundary layer plasma, a particle gyroradius $r_g$ can be considered instead of a particle mean free path $\lambda$, and cross-field radial diffusion provides a particle escape mechanism from the acceleration region. For particles with energies satisfying $r_g \ll D$, the acceleration within a finite thickness turbulent {\it shear} layer results from, both, the second order Fermi process in the turbulent medium and the {\it cosmic ray viscosity}. We consider the jet velocity to be comparable to the light velocity even on kiloparsec scales, $U \sim c$, and a characteristic velocity of magnetic turbulence, $V$, to be comparable to the Alfv\'{e}n velocity for a subsonic turbulence and at least a few orders of magnitude smaller than $U$. The Fermi process dominates over the viscous acceleration when the particle energy is small enough to satisfy $r_{g} < D \, (V / U)$. Because of rapid energy radiation losses at high energies, the electrons are always expected to satisfy the required condition. Then, the acceleration time scale in a highly turbulent shear layer can be estimated (Ostrowski 2000) as
\begin{equation}
T_{acc} \sim {r_{g} \over c} {c^{2} \over V^{2} } .
\end{equation}
With anticipated parameters of the boundary layer -- $B \approx 10^{-5} \, {\rm G}$ \ and $V \approx 2 \cdot 10^{8} \, {\rm cm/s}$ -- the characteristic acceleration time scale (1) is $T_{acc} \sim 10^2 \, \gamma \, {\rm [s]}$, where $\gamma$ is the electron Lorenz factor. One should note, that equation (1) illustrates the most optimistic scenario for the acceleration process, with presence of non-linear MHD turbulence at the scale of $r_g$. The time scale for particle escape from the acceleration region due to cross-field diffusion can be estimated as
\begin{equation}
T_{esc} \sim {D^2 \over \kappa_{\bot} } , 
\end{equation}
where $\kappa_{\bot}$ is the cross-field diffusion coefficient, $\kappa_{\bot} = \eta \, r_g \, c / 3$, and $\eta$ is a scaling numerical factor. The particle escape becomes important when the electron energy satisfies condition $T_{acc} \geq T_{esc}$ , i.e. when $r_g \geq (3 / \eta)^{1/2}\, D \, (V / c)$. For the considered boundary layer parameters and a strong turbulence condition $\eta \approx 1$, equation (2) reads as $T_{esc} \sim 6 \cdot 10^{24} \, \gamma^{-1} \, {\rm [s]}$, and hence radiative (synchrotron) cooling with a time scale $T_{cool} \sim 8 \cdot 10^{18} \, \gamma^{-1} \, {\rm [s]}$ becomes important for the much lower electrons energies. Therefore, by comparing the acceleration time scale (1) with the time scale for radiative losses, one obtains the maximum electron energy, $\gamma_{eq} \sim 10^{8}$. Rapid radiative cooling of electrons with such high maximum energy is compensated by the continuous acceleration taking place within the whole boundary layer along the jet. For lower turbulence amplitude the acceleration process can proceed slower to yield a lower value for $\gamma_{eq}$.

In the present model, after switching on the injection of seed electrons with a small initial Lorentz factor $\gamma_0$, a time dependent spectral evolution of particles accelerated in the turbulent layer begins as a power-law distribution with a growing energy cut-off. Then it evolves into a characteristic shape consisting of two different components: a stationary flat power-law part at low energies, $n_e(\gamma) \propto \gamma^{-2}$, finished with a nearly monoenergetic growing peak, or bump, at $\gamma = \gamma_{eq}$, due to an accelerated particles pile-up caused by losses (cf. Ostrowski 2000). Such electron spectrum can be approximately given by a formula using the Haeviside function, $\Theta$, and the Dirac delta function, $\delta$. At energies above the injection energy, $\gamma_0$, the electron distribution
\begin{equation}
n_{e}( \gamma, \, t) = a \, \gamma^{-2} \, \Theta( \gamma_{max}(t) - \gamma) + b(t) \, \delta( \gamma - \gamma_{eq}) \, \Theta( \gamma_{t} - \gamma_{eq}) , 
\end{equation}
where $a$ and $b(t)$ are normalization parameters (and $b(t)$ is a growing function of time), $\gamma_{t}$ is the highest energy the electrons can reach since the injection with radiation losses neglected, and $\gamma_{max}(t) = \min \left( \gamma_{t}, \, \gamma_{eq} \right)$.

\section{Radiation of electrons accelerated in a shear layer}  

We consider two radiation processes taking place at the relativistic jet boundary: the synchrotron (SYN) emission of ultrarelativistic electrons with the energy spectrum (3), and their inverse Compton (IC) cooling. Synchrotron self-absorption can be neglected with the assumed optical depth for this process to be much less than unity for $\nu \geq 10^{10} \, {\rm Hz}$. Additionally, the present preliminary IC scattering calculations are limited to the Thomson regime assuming a step scattering cross section. A high energy spectral cut-off resulting from absorption of gamma rays due to photon-photon pair creation on a cosmic infrared background is expected to occur for distant sources (e.g. Renault et al. 2001). Therefore, the obtained below photon spectra above TeV energies cannot be directly compared to observations. In the case of IC emission of a tens-of-kpc scale jets, we study the synchrotron self-Compton (SSC) radiation and Compton scattering of external CMB photons (EC). With the assumed parameters, the magnetic field energy density at the jet boundary is $u_B \approx 4 \cdot 10^{-12} \, {\rm erg/cm^3}$. The energy density of the CMB field in a source frame moving with a bulk Lorentz factor $\Gamma$ at a redshift $z$ is $u_{cmb} = u_{cmb}^{\ast} \, \Gamma^2 \, (1+z)^4$, where an asterisk denotes quantities in our local rest frame, and $u_{cmb}^{\ast} \sim 4 \cdot 10^{-13} \, {\rm erg / cm^3}$. In the jet rest frame $u_{cmb}$ can be comparable to $u_B$, and both above processes can be significant even for small $\Gamma$ and $z$.

For the optically thin emission, radiation intensities of the power-law (`1') and the high energy bump (`2') electron components (c.f. Eq.~3) are approximatly
\begin{eqnarray}
I_{syn}^{(1)}(\nu, \, t) \propto \nu^{-1/2} \, \Theta(\nu_{syn}(t) - \nu) & , & I_{syn}^{(2)}(\nu, \, t) \propto \nu^{1/3} \, \Theta(\nu_{syn}(t) - \nu) \, \Theta(\gamma_t - \gamma_{eq}) , \\
I_{ssc}^{(1+2)}(\nu, \, t) \propto \nu^{-1/2} \, \Theta(\nu_{ssc}(t) - \nu) & , & I_{ssc}^{(2)}(\nu, \, t) \propto \nu^{1/3} \, \Theta(\nu_{ssc}(t) - \nu) \, \Theta(\gamma_t - \gamma_{eq}) , \\ 
I_{ec}^{(1)}(\nu, \, t) \propto \nu^{-1/2} \, \Theta(\nu_{ec}(t) - \nu) & , & I_{ec}^{(2)}(\nu, \, t) \propto \nu^{2} \, \Theta(\nu_{ec}(t) - \nu) \, \Theta(\gamma_t - \gamma_{eq}) ,
\end{eqnarray}
where, for the assumed large scale jet parameters and neglecting the cosmological redshift correction, the maximum synchrotron, self-Compton and external Compton frequencies are $\nu_{syn}(t) \approx 10 \, \gamma_{max}^2(t) \, {\rm Hz}$, $\nu_{ssc}(t) \approx \min \left[ \nu_{syn}(t) \, , \, mc^2/h \gamma_{max}(t) \right] \, \gamma_{max}^2(t)$ and $\nu_{ec}(t) \approx 10^{11} \, \Gamma \, \gamma_{max}^2(t) \, {\rm Hz}$, respectively. In deriving equations (4) - (6) we used the standard theory of the synchrotron and the inverse-Compton radiation with simple approximations allowing evaluation of analytic expressions (e.g. Crusius \& Schlickeiser 1986 and Dermer 1995, respectively; for details see forthcoming paper Stawarz \& Ostrowski 2001). The intensities (4) - (6) allow to describe the photon spectrum variations resulting from the particle distribution evolution given in equation (3). In the model, from the beginning of injection the accelerated electrons form the power-law spectrum with the energy cut-off growing with time. Then $ \gamma_{max}(t) = \gamma_{t}$ and the resulting emission is a power-law $ \propto \nu^{-1/2}$, with a growing cut-offs $\nu_{syn}(t)$, $\nu_{ssc}(t)$ and $\nu_{ec}(t)$. When the maximum electron energy approaches $\gamma_{eq}$, the high energy particle density peak forms and starts to grow. Then, an additional radiation components $ \propto \nu^{1/3}$ (for the synchrotron radiation), $ \propto \nu^{-1/2}$ and $ \propto \nu^{1/3}$ (for the SSC emission), and $ \propto \nu^{2}$ (for the EC emission) appear and eventually dominate the spectrum at highest X-ray and $\gamma$-ray frequencies. The spectral index $\alpha = -1/3$ results from the asymptotic expansion of the synchrotron emissivity function of the `peak' electrons with the energy $\gamma_{eq}$ (c.f. Crusius \& Schlickeiser 1986), and the index $\alpha = -2$ corresponds to the Rayleigh-like approximation of the anizotropic (in the relativistic fluid rest frame) CMB radiation comptonized by the monoenergetic electrons.

Let us also briefly discuss the main implications of kinematic effects in the radiative jet output. For illustration, let us consider a long cylindrical jet of the radius $R_{j}$, moving with a constant flow Lorentz factor $\Gamma_j \approx 10$, surrounded by a boundary layer with a constant thickness $D \approx R_j \approx 1 \, {\rm kpc}$, (cf. Aloy et al. 2000), and the radial flow profile $\Gamma (r)$. In the local source rest frame, the synchrotron, SSC and EC intensities are given by expressions (4), (5), and (6), respectively, but the observed intensities are $I_{ec}^{\ast} (\nu^{\ast}, \, \theta^{\ast}) = \delta^{4+2 \alpha}(r, \, \theta^{\ast}) \, I_{ec} (\nu^{\ast})$, and $I_{syn/ssc}^{\ast} (\nu^{\ast}, \, \theta^{\ast}) = \delta ^{3+\alpha}(r, \, \theta^{\ast}) \, I_{syn/ssc}(\nu^{\ast})$, where $\alpha$ is a spectral index given in the respective expression (4) - (6), and $\delta (r, \, \theta^{\ast})$ is a Doppler factor for a sub-layer at a distance $r$ from the jet axis and the jet inclination to the line of sight equal $\theta^{\ast}$. Because of a velocity shear, the boundary layer emission beaming pattern is different compared to the uniform beaming pattern of a spine radiation. For a given viewing angle $\theta^{\ast}$, boundary layer emission comes mostly from the region with $\Gamma(r) = 1 / \sin \theta^{\ast}$, where the Doppler factor is maximized. As a result, the observed jet-counterjet brightness asymmetry is decreased in comparison to the uniform jet with $\Gamma = \Gamma_j$, allowing the spine to be more relativistic in the large scale jets than traditional estimates suggest (c.f. Komissarov 1990). 

\begin{figure}[t]
\begin{center}
\psfig{file=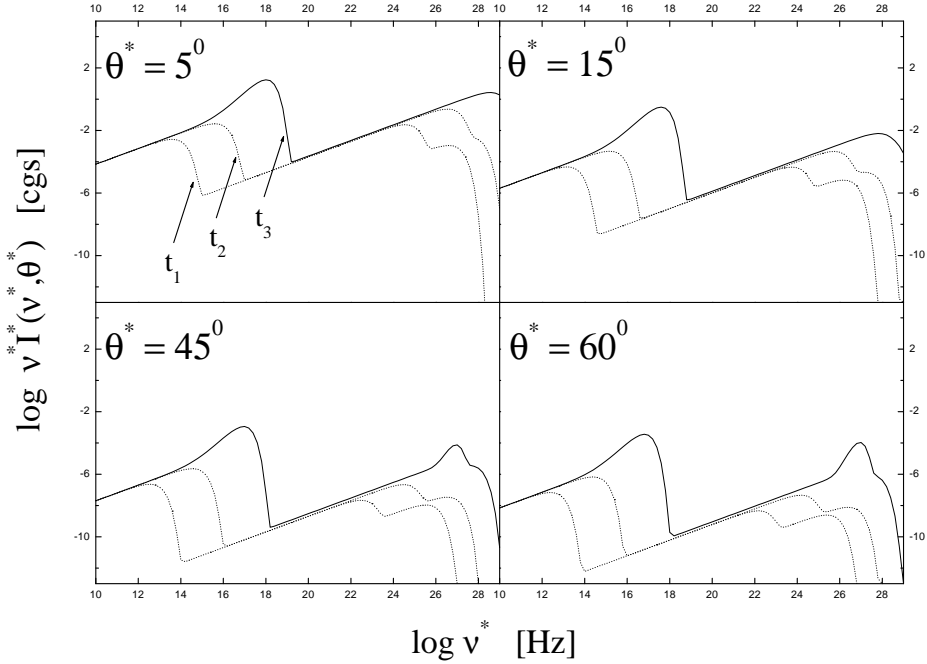,height=10cm}
\caption{{\small The observed boundary layer emission, $\nu^{\ast}\, I^{\ast}(\nu^{\ast}, \, \theta^{\ast})$, for four different viewing angles $\theta^{\ast}$, as measured near the jet, $z=0$. Lines $t_1$, $t_2$, and $t_3$ correspond to $\gamma_{max}(t) = 10^6$, $\gamma_{max}(t) = 10^7$, and $\gamma_{max}(t) = \gamma_{eq}=10^8$, respectively. }  }
\end{center}         
\end{figure}

The observed boundary layer emission, $\nu^{\ast} \, I^{\ast}(\nu^{\ast}, \, \theta^{\ast})$, are plotted on Fig.~1 for four different viewing angles. Lines indicated by $t_1$, $t_2$ and $t_3$ correspond to different forms of the particle spectrum obtained in the considered acceleration model in successive times from the beginning of the acceleration process. For instance the line $t_1$ illustrates resultant sum of synchrotron, self-Compton and external Compton emission of the power-law electron component (index `1' in Eq.~4 - 6) with an energy cut-off $\gamma_{max}(t) = 10^6 < \gamma_{eq}$, while the line $t_3$ includes also the additional contribution from the high energy electron bump at $\gamma_{eq} = 10^8$ (index `2' in Eq.~4 - 6). The presented spectra are integrated over the line of sight with the assumed linear dependence $\Gamma = \Gamma(r)$ within the shear layer. 

\section{Conclusions}

In the present paper, we study radiation of ultrarelativistic electrons accelerated at the boundary shear layer of the large scale relativistic jet. The considered acceleration process generate a characteristic particle distribution consisting of two components: the power-law section extended up to a more or less narrow pile-up bump, modeled here as the monoenergetic peak. A possible characteristic of such a cosmic ray population is its time variability, constrained by the not considered here, non-linear particle backreaction at the acceleration process. 

The acceleration within the jet boundary layer with inefficient particle escape due to radial diffusion could increas the energetic particle energy density $u_e$ to extremely high values. However, when $u_e$ becomes comparable to the ambient medium energy density, the accelerated particles can influence physical conditions within the boundary region. The involved electron pressure tends to expand the acceleration region rarefying the medium and, possibly, decreasing the efficiency of the acceleration process (c.f. Arav \& Begelman 1992, Ostrowski 2000). Changes of the boundary layer density can result in changing several characteristic of the acceleration process, in particular shifting of the local value of $\gamma_{eq}$ and changing of the flow profile at the boundary. Such particle pressure backreaction, leading to local temporal variations of the high energy electron component, will generate fluctuations of the high energy radiation over the stationary power-law component emission at low photon energies. For the assumed physical conditions with $\gamma_{eq} = 10^8$, the process involves variations at X-ray and $\gamma$-ray bands with energies of $\sim$ keV and $\sim$ TeV, respectively. In our study of the tens-of-kpc scale jets, the involved time scales are $\sim 10^{10} \, {\rm s}$, and, thus, can result in spatial kpc-scale brightness variations along the jet boundary. Much shorter scales are possibly involved in (sub) mili-arc-second jets, with much smaller (a factor of $10^3$) spatial scales, much larger (a factor of $10^5$) magnetic fields, and stronger external radiation fields (study in progress now). 

As the described above radiative spectral components appear in a natural way in the discussed process, the stationary or time averaged spectra possibly formed due to the mentioned non-linear effects are also expected to posses such components. Then, one should observe X-ray and $\gamma$-ray excesses in electromagnetic spectrum of the jet, as illustrated on Fig.~1. With the jet parameters considered above, such an X-ray flux above the extrapolated radio-to-optical continuum can explain the keV emission detected in several jets by Chandra.

One should note, that our simple model, far from being complete, is supposed to describe main radiative properties of the boundary region under simple assumptions, and to provide restrictions on physical conditions within such layer.

\section*{Acknowledgements}
We are grateful to Marek Sikora for his help and discussions. The present work was supported by Komitet Bada\'{n} Naukowych through the grant BP 258/P03/99/17. 

\section*{References}

\reference Aloy, M.A., Ibanez, J.M., Marti, J.M., Gomez, J.L., Muller, E. 1999, ApJ, 523, L125
\reference Aloy, M.A., Gomez, J.L., Ibanez, J.M., Marti, J.M., Muller, E. 2000, ApJ, 528, L85
\reference Arav, N., Begelman, M.C. 1992, ApJ, 401, 125 
\reference Attridge, J.M., Roberts, D.H., Wardle, J.F.C. 1999, ApJ, 518, L87
\reference Berezhko, E.G. 1990, Preprint {\it Frictional Acceleration of Cosmic Rays}, The Yakut Scientific Centre, Yakutsk
\reference Crusius, A., Schilckeiser, R. 1986, A\&A, 164, L16
\reference Dermer, C.D. 1995, ApJ, 446, L63
\reference Earl, J.A., Jokipii, J.R., Morfill, G. 1988, ApJ, 331, L91 
\reference Edwards, P.G., Giovannini, G., Cotton, W.D., Feretti, L., Fujisawa, K., Hirabayashi, H., Lara, L., Venturi, T. 2000, PASJ, 52, 1015
\reference Jester, S., Roser, H.J., Meisenheimer, K., Perley, R., Conway, R. 2001, A\&A, 373, 447
\reference Jokipii, J.R., Kota, J., Morfill, G. 1989, ApJ, 345, L67
\reference Komissarov, S.S. 1990, SvAL, 16, 284 
\reference Ostrowski, M. 1990, A\&A, 238, 435 
\reference Ostrowski, M. 1998, A\&A, 335, 134 
\reference Ostrowski, M. 2000, MNRAS, 312, 579
\reference Ostrowski, M., Sikora, M. 2001, in {\it Proc. 20th Texas Symp. on Relativistic Astrophysics}, eds. J.C. Wheeler \& H. Martel, Austin.
\reference Owen, F.N., Hardee, P.E., Cornwell, T.J. 1989, ApJ 340, 698
\reference Perlman, E., Biretta, J., Fang, Z., Sparks, W., Macchetto, F. 1999, AJ, 117, 2185
\reference Renault, C., Barrau, A., Lagache, G., Puget, J.-L. 2001, A\&A, 371, 771
\reference Stawarz, \L ., Ostrowski, M. 2001, {\it in preparation}
\reference Swain, M.R., Bridle, A.H., Baum, S.A. 1998, ApJ, 507, L29

\end{document}